\documentclass[preprintnumbers,amsmath,11pt,amssymb,floatfix,superscriptaddress,nofootinbib]{article}

\def\mytitle#1{\setcounter{equation}{0}
\setcounter{footnote}{0}
\begin{flushleft}\Large\textbf{#1}\end{flushleft}
\vspace{0.25cm}}
\def\myname#1{\leftline{{\large #1}}\vspace{-0.13cm}}
\def\myplace#1#2{\small\begin{flushleft}\textit{#1}\\
\texttt{#2}\end{flushleft}}

\def\myclassification#1{\small\noindent
Keywords :
       #1\vspace{0.5cm}}
\usepackage{graphicx}
\begin{document}

\mytitle{Thermodynamic Products for Einstein-Gauss-Bonnet Black Hole with $\alpha$-Corrected Entropy Term}

\myname{$Abhijit~ Mandal^{*}$\footnote{abhijitmandal.math@gmail.com}, $Ritabrata~
Biswas^{\dag}$\footnote{biswas.ritabrata@gmail.com}}
\myplace{*Department of Mathematics, Jadavpur University, Kolkata $-$ 700 032, India.\\$\dag$ Department of Mathematics, Bankura University, Bankura $-$ 722 146, India.}{} 
 
\begin{abstract}
In the present work, we consider a charged black hole in five dimensional Einstein-Gauss-Bonnet gravity where the $\alpha$ corrected entropy term is considered. We examine the horizon radii  product, entropy product, Hawking temperature product and free energy product for both event horizon and Cauchy horizon. Our motive is to check whether the same quantity for event horizon and Cauchy horizon is free of mass, i.e., global or not. We further study the stability of such black hole by computing the specific heat and free energy for both the horizons. All these calculation might be helpful to understand the microscopic nature of such black holes.
\end{abstract}

\myclassification{Black hole physics; Thermodynamic process; Relativity and gravitation.}

Low energy action of string theory and quantum field theory for curved space time gives rise of higher derivative curvature terms\cite{Maldacena}. AdS-CFT corrospondence\cite{Klebanov} constructs corrections of large $N$ expansion of boundary CFTs on the side of dual field theory which is seen to be the higher derivative curvature terms. Due to the nonlinearity of Einstein's equations, it is difficult to find exact solutions. To do this, approximations are highly required.

Gauss Bonnet Lagrangian does consider upto quadratic curvature as 
$${\cal L}_{GB} = R_{\mu \nu \gamma \delta} R^{\mu \nu \gamma \delta} - 4 R_{\mu \nu} R^{\mu \nu} + R^{2} $$
The Einstein gravity modified with Gauss Bonnet term constructs equation of motion having no more than second derivatives of metric and the theory has been shown to be free of ghosts when it is expanded about flat space, evading any problems with unitary\cite{Witten1}. Also Gauss Bonnet term acts as the leading correction\cite{Witten2} to the effective low energy action of the string theory. Thermodynamic aspects of Gauss Bonnet black hole(BH hereafter) in asymptotically flat space have been analysed in Refs like \cite{Huang, Cai2}. 

Gauss Bonnet BHs in asymptotically de-Sitter space was studied by Cai\cite{Cai1}. Due to the absence of spatial infinity and a globally timelike Killing vector in dS space, calculation of the conserved charges associated with an asymptotically dS space is bit non-trivial. To construct our entropy candidate for this Gauss-Bonnet BHS, we use the same thermodynamic procedure which are used in Einstein gravity. Generally, in higher derivative gravity theories curvature terms, entropy of the BHs doesnot satisfy the area formula.

	 From \cite{Lemos, Huang}, the entropy of the Gauss Bonnet BH in dS space reads as
$$S=\frac{\Omega_{d-2} r_+^{d-2}}{4G} \left\{1+\frac{2(d-2)\tilde{\alpha}}{(d-4)r_+^2}\right\}.$$ 

For $\alpha > 0$, this expression is manifestly non-negative and for $\alpha < 0$, the entropy again always increasesbut can be negative for small $r_+$. The minimum entropy occurs as $r_+ \rightarrow |\alpha|$, at which point the horizon vanishes exposing a naked singularity.

The action of the Einstein Gauss Bonnet(EGB hereafter) Gravity in $5$ dimensional space time ($M,g_{ij}$) can be written as \cite{Cai} (Taking $8\pi G = c = 1$ as unit)
\begin{equation}
S = \frac{1}{2}\int_{M} d^5 x\sqrt{-g}[R + \alpha R_{GB} + L_{matter}],
\end{equation}
$\alpha$ is the coupling constant of the GB term having dimension $(length)^2 (\alpha \geq 0)$.
 
  $L_{matter} = F_{\mu\nu} F^{\mu\nu}$ is the matter Lagrangian where $F_{\mu\nu} = \partial_{\mu} A_{\nu} - \partial_{\nu} A_{\mu}$ is the electromagnetic tensor field, $A_{\mu}$ is the vector potential. 
	The gravitational and electromagnetic field equations obtained by varying the
action (i.e., $\delta S = 0$) with respect to $g_{\mu\nu}$ and $F_{\mu\nu}$ are
\begin{equation}\label{EGB}
G_{\mu\nu}= T_{\mu\nu}^{EM} + T_{\mu\nu}^{GB}
\end{equation}
and
\begin{equation}\label{Maxweel_equation}
\nabla_{\mu} F_{\nu}^{\mu} = 0,
\end{equation}
where, $T_{\mu\nu} = \alpha H_{\mu\nu}$, where $H_{\mu\nu}$ is the Lovelock tensor given by,
\begin{equation}
H_{\mu\nu} = 2[R R_{\mu\nu} - 2R_{\lambda\nu}R_{\nu}^{\lambda} - 2R^{\gamma\delta}R_{\mu\gamma\nu\delta} + R_{\mu}^{\alpha\beta\gamma}R_{\nu\alpha\beta\gamma}] - \frac{1}{2} g_{\mu\nu}R_{GB}
\end{equation}
and $T_{\mu\nu} = 2F_{\mu}^{\lambda} F_{\lambda\nu} − \frac{1}{2} F_{\lambda\sigma} F^{\lambda\sigma} g_{\mu\nu}$ is the electromagnetic stress tensor.

The Gauss Bonnet Lagrangian $R_{GB}$ is only non-trivial in $(4+1)D$ or greater, and as such, only applies to extra dimensional models. In $(3+1)D$ and lower it reduces to a topological surface term.

We now proceed to solve the field equations (\ref{EGB}) for the five-dimensional static spherically symmetric space time with the line element
\begin{equation}
ds^2 = -f(r) dt^2 + \frac{dr^2}{f(r)} + r^2 d\Omega ^2_3
\end{equation}
where $d\Omega ^2_3$ is the metric of a $3D$ hyper surface with the constant curvature $6K$ having an explicit form
\begin{equation}
d\Omega ^2_3 = 
\left\{
\begin{array}{lll} 
d\theta_1^2 + sin^2 \theta_1 (d\theta_2^2 + sin^2 \theta_2 d\theta_3^2 ) , \mbox{(K= 1)}\\\\
d\theta_1^2 + sin^2 \theta_1 (d\theta_2^2 + sinh^2 \theta_2 d\theta_3^2 ), \mbox{(K= -1)}\\\\
\alpha^{-1} dx^2 + \sum^2_{i=1} d\phi^2_i,~~~~~~~~~ \mbox{(K=0)}
\end{array}
\right.
\end{equation}
where the coordinate $x$ has the dimension of length while the angular coordinates $(\theta_1, \theta_2, \theta_3)$ and $(\phi_1 , \theta_2)$ are dimensionless, with ranges
$\theta_1, \theta_2 : ~~~[0, \pi]~~~~~~~~\theta_3, \phi_1 , \phi_2 : ~~~[0, 2\pi].$

If we assume that there exists a charge $q$ at $r = 0$ (note that $q$ is a point charge for $K = \pm 1$ and is the charge density of a line charge for $K = 0$), then the vector potential may be chosen to be
$A_\mu = \phi(r)\delta_\mu^0.$
Now using the Maxwell equation (\ref{Maxweel_equation}), the differential equation for $\phi(r)$ becomes
\begin{equation}
r\frac{d^2\phi}{dr^2} + 3\frac{d\phi}{dr} = 0 \Rightarrow \phi(r) = - \frac{q}{2r^2},
\end{equation}
where the Gauss law has been used to determine the integration constant. Now the
metric function for EGB gravity, say $f_{EGB} (r)$ can be obtained by solving the field equations (\ref{EGB}) as \cite{Dehghani}
\begin{equation}\label{BH_solution}
f_{EGB} (r) = K +\frac{r^2}{4\alpha}\left[1 \pm \sqrt{1+ \frac{8 \alpha(M +2\alpha|K|)}{r^4}-\frac{8\alpha q^2}{3r^6}}\right].
\end{equation}

For BH horizons, we will make $f_{EGB} (r)=0$ to have,
\begin{equation}
r_{\pm}= \frac{1}{2}\left[\sqrt{M+\frac{2q}{\sqrt{3}}} \pm \sqrt{M-\frac{2q}{\sqrt{3}}} \right],
\end{equation}
where this $+$ sign corresponds to event horizon and the $-$ sign corresponds to Cauchy horizon. This is to be followed that the radii of horizons are independent of the coupling parameter $\alpha$.

The purpose of this work is to discussed the thermodynamic products of a charged BH solution in account of a explicit form of the thermodynamic entropy. It will be interesting to observe where the thermodynamic parameters of the event and Cauchy horizons are interacting or not.

Next, we will calculate different thermodynamic products and analyse the stability of such BHs. At the end we will conclude of this work.

Hawking's idea towards BH temperature leads to the thinking about the size of the underlying space of quantum states in quantitative detail \cite{Hawking}. These ideas are obeying the fact that BH geometry is closed inside the outer event horizon. Other geometric properties giving similarly direct evidence on the microscopic structure of BHs. One characteristic property of string models say that the entropy is the sum of contributions from left and right moving excitations of the string and the thermodynamic variables accordingly appear in duplicate versions. BH geometry follows the same. All the thermodynamic variables, defined at the outer event horizon is been copied by another independent set of thermodynamic variables defined at the inner event horizon. The left and right moving thermodynamics of string theory have a corrospondence to the sum and different of the outer and inner horizon thermodynamics\cite{Cevtic1}. The structure of the entropy as a sum of two terms may be an indication that all BHs can be described in this way and that the two terms in the entropy are the contributions from left and right moving modes. If this is true then interactions between the two kinds of modes can be treated as weak. Colliding left and right modes give rise to Hawking Radiation. Many references \cite{Cevtic2, Larsen1} shows that spacetime geometry devides the entropy and the temperature in very same way that the microscopic interpretation does.



Now, the product of both radii of event and Cauchy horizons is
\begin{equation}
r_+ r_- = \frac{q}{\sqrt{3}}, 
\end{equation}
clearly stating the fact that this quantity is universal.

Usually entropy of a BH calculated from the so-called area formula and which equals to one-quarter of the horizon area. In higher derivative curvature terms, in general, the entropy of a BH doesnot match with the area formula. But as a thermodynamic system BH must obey the first law of thermodynamics $dM = T dS$ \cite{Cai1}. Integrating the first law, we have,
$$S = \int T dM$$
and $$S_\pm = \int_0^{r_\pm} T^{-1} \frac{\delta M}{\delta r_{\pm}} dr_{\pm}$$
where we have imposed the physical assumption that entropy vanishes when the horizons of BH shrinks. This specification of lower limit gives a formula
\begin{equation}
S_{\pm} = r^3_{\pm} + 6 \alpha r_{\pm}~~~~\Rightarrow~S_+S_- = \frac{q}{\sqrt{3}}\left[ \frac{q^2}{3} + 36 \alpha^2 + 6 \alpha M\right]
\end{equation}

\begin{equation}
T_{\pm}=\frac{\partial M}{\partial S_{h/c}} = \frac{1}{\frac{\partial S_{h/c}}{\partial r_{h/c}}\frac{\partial r_{h/c}}{\partial M}} = \frac{6 r^4_{\pm} - 2 q^2}{9 r_{\pm}^3 (r_{\pm}^2 + 2 \alpha)}~~~~~and~we~have~,~
\end{equation}
\begin{equation}
T_+T_- = \frac{4 \left(4 q^2 - 3 M^2 \right)}{3 \sqrt{3} q \left( q^2 + 12 \alpha^2 + 6 \alpha M\right)}.
\end{equation}

Now, the free energy will take the form
\begin{equation}
F_{\pm}= M - T_{\pm}S_{\pm}= M - \frac{6 r^6_{\pm} + 36 \alpha r^4_{\pm} -2 q^2 r^2_{\pm} -12 \alpha q^2}{ 9 r^4_{\pm} + 18 \alpha r^2_{\pm}}
\end{equation}
The product of these two is
\begin{equation}
F_+F_- = \frac{162 \alpha M^2\left(M - 6 \alpha\right) + 3 q^2\left(5 M^2 + 576 \alpha^2 \right) + 16 q^4}{27 \left(12 \alpha^2 + 6 \alpha M + q^2\right)}
\end{equation}

\begin{equation}
C_+C_- = \pm \frac{9 \sqrt{M^2 - \frac{4 q^2}{3}}\left[\sqrt{M+\frac{2q}{\sqrt{3}}} \pm \sqrt{M-\frac{2q}{\sqrt{3}}} \right]^3\left(12 \alpha + 3M \pm 3\sqrt{M^2 - \frac{4 q^2}{3}} \right)}{144\left[ -3 M^3 + 6 \alpha M^2 + 8 q^2 M + 8 \alpha q^2 \mp \sqrt{M^2 - \frac{4 q^2}{3}}(3M^2 - 6 \alpha M -6q^2)\right]}
\end{equation}

\begin{equation}
C_+C_- = \frac{\left(6 \alpha(M+ 2 \alpha) + q^2\right)^2\left(4 q^3 - 3 M^2 q\right)}{\sqrt{3} \left( 18 \alpha^2 M(2 \alpha -M) + q^2 \left(16 \alpha^2 + 56 \alpha M -5 M^2 + 12 q^2\right)\right)}
\end{equation}

\begin{figure}[h]
\begin{center}

~~~~~~~~~~Fig.1a~~~~~~~~~~~~~~~~~~~~~~~~~~~Fig.1b~~~~~\\
\includegraphics[height=2.2in, width=1.5in]{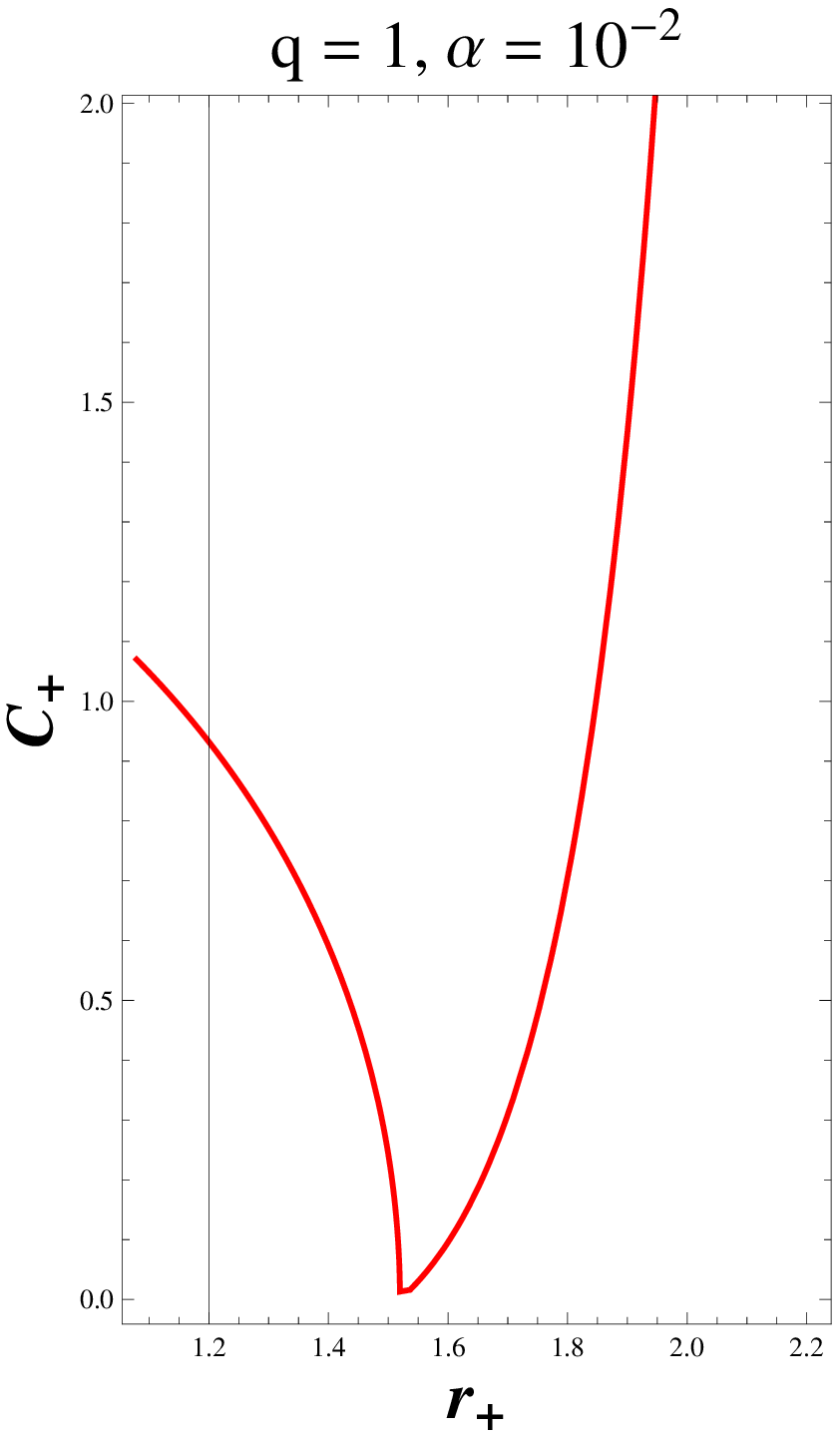}~~~~
\vspace{.5cm}
\includegraphics[height=2.2in, width=1.5in]{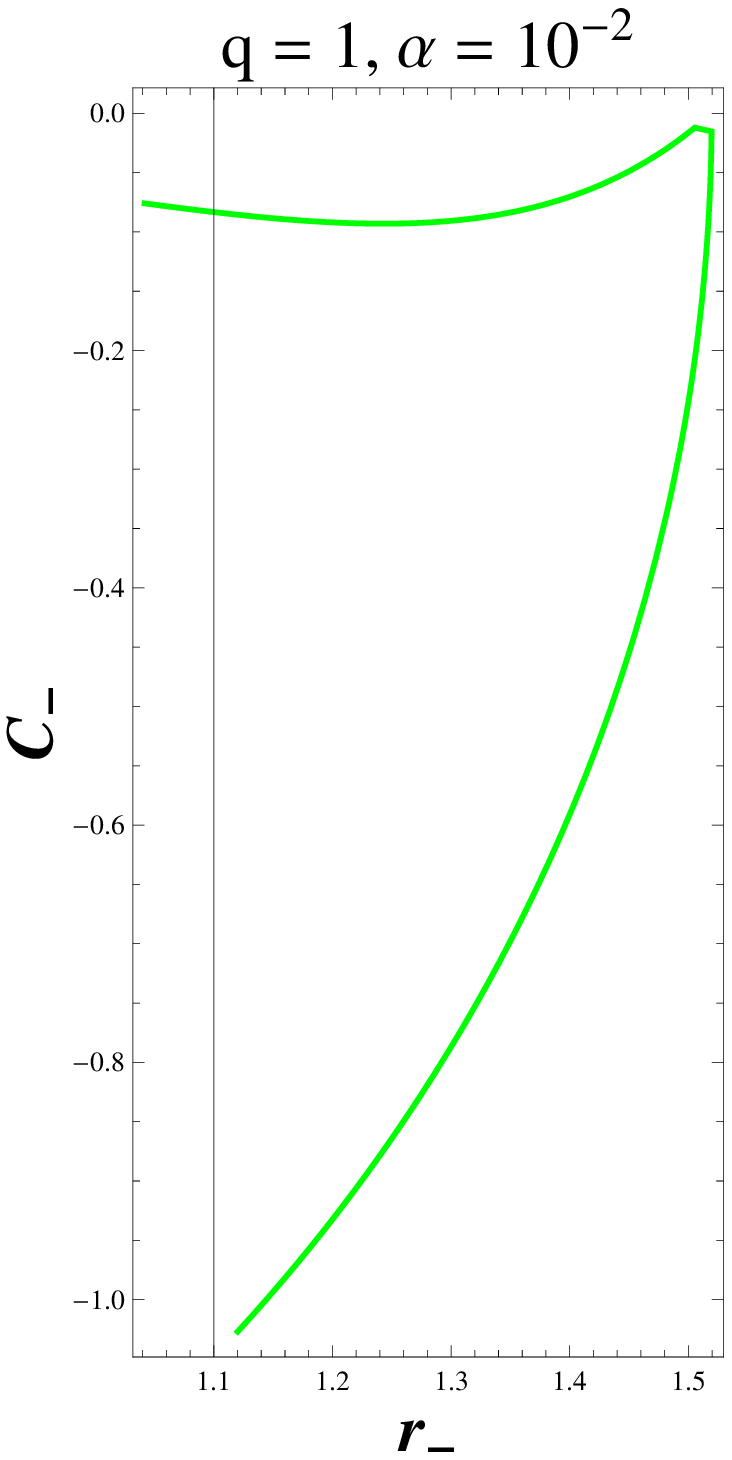}~~\\
Fig $1a$ represent the variation of specific heat($C_+$) for event horizon with respect to $r_+$ for charge $q=1$ and coupling parameter $\alpha = 10^-2$.
Fig $1b$ represent the variation of specific heat($C_-$) for Cauchy horizon with respect to $r_-$ for charge $q= 1$ and coupling parameter $\alpha = 10^-2$.

\end{center}
\end{figure}
In figure $1a$ we show the specific heat($C_+$) for event horizon against the radius of event horizon($r_+$), where charge $q=1$ and the coupling constant $\alpha = 10^{-2}$. Here, we observe two phases, one for small values of $r_+$ and another for large values of $r_+$. The $C_+$ vs $r_+$ graph for the first phase is slowly decreasing finally becomes $C_+ =0$ for a particular value of $r_+$(say, $r^{\dag}$). On the other side, the $C_+$ vs $r_+$ curve for the second phase starts from $C_+ =0$ at $r_+ = r^{\dag}$ and the curve is strictly increasing. So, both the phases have a common point at $r_+ = r^{\dag}$, which takes the value $C_+ =0$. Here, the first phase radiates energy/heat to decrease the temperature while the second phase absorb more and more energy for the increment of temperature. Over all the sign of $C_+$ is positive, which indicates that the spacetime confined within event horizon is stable.

Now, the figure $1b$ shows the variation of the specific heat($C_-$) for Cauchy horizon with respect to the radius of Cauchy horizon($r_-$), where $q=1$ and $\alpha = 10^{-2}$. In this figure, we find two phases and for each value of $r_-$, we will find two values of $C_-$. In one phase, the $C_-$ vs $r_-$ curve is slowly increasing and at last takes the value $C_- = 0$ at $r_- = r^{\dag}$. The $C_-$ vs $r_-$ is also increasing for the another phase and finally it will also becomes $C_- = 0$ at $r_-= r^{\dag}$. Here, overall the sign of $C_-$ is negative, which indicating that the spacetime confined within Cauchy horizon is unstable. But, for $r_- > r^{\dag}$ we find no value/ physical curve of $C_-$, which may be indicates that the Cauchy horizon does not exist after that.

\begin{figure}[h]
\begin{center}

~~~~~~~~~~Fig.2a~~~~~~~~~~~~~~~~~~~~~~~~~~~Fig.2b~~~~~\\
\includegraphics[height=1.5in, width=2.2in]{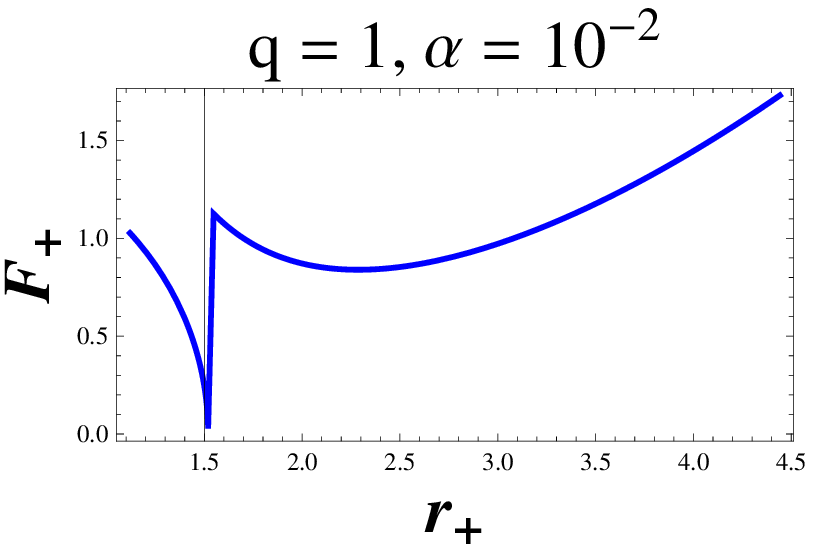}~~~~
\vspace{.1cm}
\includegraphics[height=1.5in, width=2in]{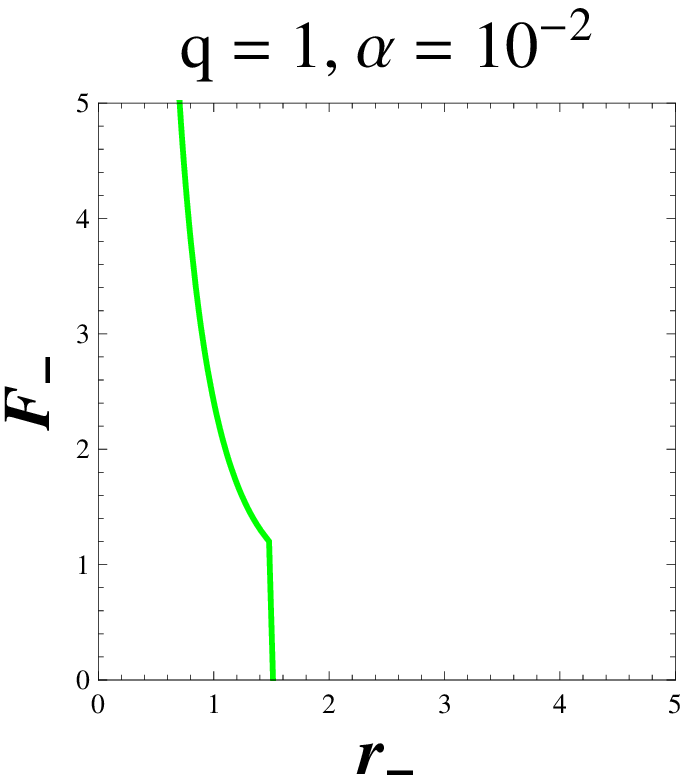}~~\\
Fig $2a$ represent the variation of free energy($F_+$) for event horizon with respect to $r_+$ for charge $q=1$ and coupling parameter $\alpha = 10^-2$.
Fig $2b$ represent the variation of free energy($F_-$) for Cauchy horizon with respect to $r_-$ for charge $q= 1$ and coupling parameter $\alpha = 10^-2$.
 \end{center}
\end{figure}

We analyze the free energy($F_+$) for event horizon in figure $2a$, in which $ F_+ $ is plotted with respect to the radius of event horizon($r_+$) with $q=1$ and $\alpha=10^{-2}$. In this figure when the value of $r_+$ is small($r_+ < r^{\dag} $), we will find a phase in which the $F_+$ and $r_+$ curve is strictly decreasing and when the value of $r_-$ is large($r_+ > r^{\dag} $), we observe a phase in which the curve is increasing. The first phase ends at $r_+ = r^{\dag}$ and the second phase starts at $r_+ = r^{\dag}$. Here, we will find a finite energy jump at $r_+ = r^{\dag}$. Since, the specific heat($C_+$) for event horizon vanishes at $r_+= r^*$, may be there is a huge amount of energy absorption happens which is the reason behind this energy jump at $r_+= r^{\dag}$. Overall the sign of $F_+$ is positive which signifies the stability within event horizon.


Now to conclude we should mention at the very first that only the BH radii product is universal quantity, whereas the Hawking temperature product, entropy product, free energy product and specific heat product are not universal quantities because they all are depends on mass parameter. We also studied the stability of event and Cauchy horizon from the $C_+ ~vs~ r_+$ and $C_-~ vs~ r_-$ curves. The spacetime confined with in event horizon is stable but if we follow the space time confined inside CH, this is always unstable.

Here, between the radius of event horizon $r_+$ and the radius of Cauchy horizon($r_-$), the co-ordinates $r$ becomes timelike. Even we found two different phases by analyzing the specific heat and the free energy at the Cauchy horizon. This two phases exist for the same value of $r$ though and at a particlar radius of Cauchy horizon they meet each other. At this common cuspidal type meeting point of these two phases the specific heat is zero. This means even not absorbing or radiating, the BH can increase its temperature at this particular critical radii. Once $r>r_-$ we find no value/ physical curve of $C_-$, which indicating the fact that the Cauchy horizon for such a big BH even does not exist at all.

\vspace{.1 in}
{\bf Acknowledgement:}
RB thanks Inter University Center for Astronomy and Astrophysics(IUCAA), Pune, India for Visiting Associateship. Authors thank IUCAA for local hospitality. This work was done during a visit there. Authors thank Prof. Subenoy Chakraborty, Department of Mathematics, Jadavpur University for fruitful discussions.

\end{document}